\documentclass{INTERSPEECH2023}
\usepackage{amsmath}
\usepackage{color, colortbl}
\usepackage{array,multirow,graphicx}
\usepackage{tikz}
\usetikzlibrary{decorations.pathreplacing}

\usepackage[belowskip=-8pt,aboveskip=5pt]{caption}
\usepackage{float}

\interspeechcameraready

\title{Zero Shot Text to Speech Augmentation for Automatic Speech Recognition on Low-Resource Accented Speech Corpora}
\name{Francesco Nespoli$^{1,2}$, Daniel Barreda$^1$, Patrick A. Naylor$^2$}
\address{
  $^1$Microsoft \\
  $^2$Imperial College London}
\email{fnespoli@microsoft.com, daniel.almendrobarreda@microsoft.com, p.naylor@imperial.ac.uk}

\let\OLDthebibliography\thebibliography
\renewcommand\thebibliography[1]{
  \OLDthebibliography{#1}
  \setlength{\parskip}{0pt}
  \setlength{\itemsep}{1.8pt plus 1ex}
}
\begin{document}

\maketitle
 
\begin{abstract}
    In recent years, automatic speech recognition (ASR) models greatly improved transcription performance both in clean, low noise, acoustic conditions and in reverberant environments. However, all these systems rely on the availability of hundreds of hours of labelled training data in specific acoustic conditions. When such a training dataset is not available, the performance of the system is heavily impacted. For example, this happens when a specific acoustic environment or a particular population of speakers is under-represented in the training dataset. Specifically, in this paper we investigate the effect of accented speech data on an off-the-shelf ASR system. Furthermore, we suggest a strategy based on zero-shot text-to-speech to augment the accented speech corpora. We show that this augmentation method is able to mitigate the loss in performance of the ASR system on accented data up to 5\% word error rate reduction (WERR). In conclusion, we demonstrate that by incorporating a modest fraction of real with synthetically generated data, the ASR system exhibits superior performance compared to a model trained exclusively on authentic accented speech with up to 14\% WERR.  
    
\end{abstract}
\noindent\textbf{Index Terms}:  speech recognition, data augmentation, text to speech, accented data

\section{Introduction}
    Contemporary ASR systems necessitate a substantial volume of meticulously labeled speech data to yield accurate and effective transcriptions. Many of these sophisticated models leverage open-source datasets such as Librispeech [1], as integral components of their training. However, the transition from controlled training environments to real-world applications results in a notable disparity in performance. This discrepancy can be attributed to a superposition of different factors, including diverse acoustic conditions and the under-representation of certain language groups within the training data. While it is widely acknowledged that the mere presence of bias in training data does not inherently translate to biased models [2], the subtle speech variations introduced by regional accents among native speakers [3] and even racial characteristics [4] pose potential sources of bias in ASR models. The substantial decline in ASR system performance when confronted with non-native speech patterns provides empirical evidence that models primarily trained on the utterances of native speakers lack the required robustness to effectively model underrepresented pronunciation patterns. These findings underline the importance for a more inclusive approach in the development and training of ASR systems. \\ Many different techniques have been suggested to mitigate ASR degradation. In the context of English accented speech data, the authors in \cite{li2021accent} provide the ASR model an extra input carrying accent information. In this case, the extra input is an accent embedding which is computed from an external pre-trained model. This strategy has been shown to be beneficial for the ASR system in the form of an improved robustness to different accents. Another effective technique is the application of domain adversarial neural networks (DANN). This technique employs a domain discriminator (DD) in addition to the primary task network (ASR in this case). The idea behind DANN is to push the ASR front-end to extract domain-invariant representations. During training, the DD tries to distinguish between the source and target domains based on the feature representations extracted by the model, while the primary task network aims to perform its main task. However, at the optimization stage, the DD loss is subtracted from the ASR loss effectively pushing the feature extractor to learn domain-invariant representations. When applied on accented speech, domain adversarial training has been shown to diminish the mismatch between the accented and standard speech distributions (at the feature level) therefore leading to a more robust ASR model for accented speech \cite{DNN}. Finally, accent conversion techniques have been investigated for converting foreign to native \cite{f2n} and native to foreign accents \cite{n2f}. In the first case, foreign speech is normalized to the native accent and then the resulting audio signal is fed to the ASR model. In the second case, native speech can be employed to synthetically produce accented utterances. These data can then be mixed with the original training corpus to perform data augmentation. \\ In this work, we investigated the impact of accented data on ASR systems performance. Our findings underscore the need for innovative strategies to enhance ASR model performance in presence of linguistic variations and accents. To address this challenge, we propose the integration of a state-of-the-art zero-shot text-to-speech (ZS-TTS) system for the augmentation of accented data. The rationale behind this lies in the inherent capability of ZS-TTS to  assimilate the acoustic characteristic of a specific speaker, by utilizing only few seconds of training data. These features prove particularly advantageous in the context of accented corpora, where datasets predominantly consist of a limited number of speakers, coupled with a scarcity of recordings. Moreover, the distinguishing feature of ZS-TTS lies in its capacity for virtually limitless data production from each available speaker. In the context of accented corpora, this unique attribute can be strategically harnessed to significantly augment the training dataset for the ASR model, thereby effectively mitigating the inherent mismatch between the training and production data. \\ Our methodological approach involves the utilization of the ZS-TTS system as outlined in \cite{yourtts}, coupled with the employment of a transformer-based ASR model from \cite{speechbrain}. This choice is motivated by the openness and ease of training and deployment of both models. Furthermore, our selection is underscored by the multi-lingual capabilities embedded in ZS-TTS, a feature that we have repurposed and exploited to address the challenges posed by multi-accent learning.

\section{Proposed Method}

\subsection{Zero Shot Text-to-Speech}

    We exploit YourTTS \cite{yourtts}, a multi-speaker and multilingual TTS architecture employing variational inference and adversarial learning \cite{vits}. 
    \begin{figure}[ht]
    \centerline{\includegraphics[scale=0.3]{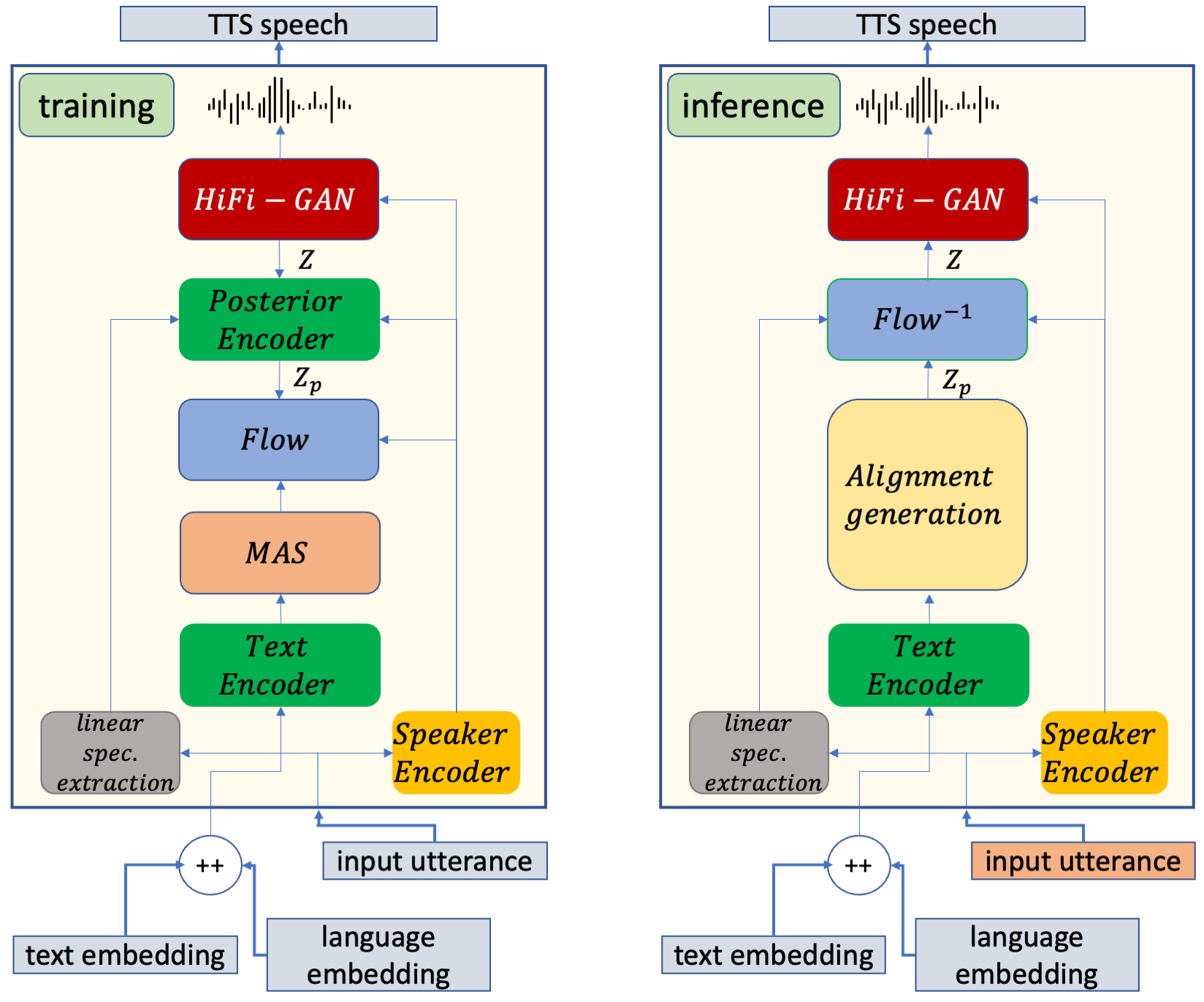}}
     \caption{Overviwew of the Zero Shot TTS architecture. Left: training stage. Right: inference stage.}
    \label{fig1}
    \end{figure} The model has five main components: text encoder, transform encoder, alignment stage, an invertible flow-based decoder and a vocoder \cite{hifi}. While training, we input the linear spectrogram to the posterior encoder and it outputs a latent representation $Z$. This goes into the vocoder and the flow-based decoder whose output $Z\textsubscript{p}$ is then aligned with the text encoder representation with monotonic alignment search (MAS) \cite{glowtts}. At inference time, the text encoder generates the alignment $Z\textsubscript{p}$ which is then used by the inverted flow decoder to produce $Z$ which, when input into the vocoder, generates an audio. The speaker encoder in Fig.~1 is the H/ASP architecture \cite{HASP} based on ResNet-34 and generates embeddings used for the speaker conditioning task. The model is also able to learn multiple languages. This objective is accomplished through the utilization of trainable language embeddings, as illustrated in Fig.~1. These vectors, characterized by four dimensions, are appended to the embeddings associated with each input character, thereby generating a language-specific character representation. Expanding on this concept, in instances where multiple accents are present, we leverage the language embedding as an accent embedding. Initially, we start with an open-source multi-lingual checkpoint that encompasses French, Portuguese, and English \cite{yourtts}. Subsequently, we undertake the random initialization of the existing $N$ language embedding layers. If $M$ denotes the number of accents within the training data, we introduce an additional set of random accent embeddings, totaling $|M-N|$, to accommodate every accent featured in the training set. This augmentation ensures a comprehensive coverage of accent variations, enhancing the model's ability to capture and represent diverse linguistic nuances. Moreover, the integration of accent embeddings enriches the accent-dependent character representations, contributing to a more contextually aware language model.

\subsection{Automatic Speech Recognition Model}

     The experimental results shown in Section~3 have been computed by employing an ASR system based on a transformer acoustic model encoder and a joint transformer decoder combining connectionist temporal classification (CTC) \cite{CTC}, with decoding stage integrating also CTC probabilities. The ASR model was implemented with \cite{speechbrain} and was pre-trained for 90 epochs on Librispeech 960 hours with a batch size of 16, gradient accumulation factor 16, Adam optimizer with learning rate 0.001 and noam learning rate decay scheme.
    
\section{Experimental Data and Results}

    We conducted both single and multi accent experiments. In the first case, we used the VCTK \cite{vctk} dataset from which we selected 3 English speakers with Indian accent. In our multi-accent experiments, we used the Interspeech 2020 accented english recognition challenge dataset \cite{shi2021accented}. The dataset contains ten distinct subsets of accented English recordings     namely: United States (USA), Canadian (CAN), Great Britain (GBR), Chinese (CHN), Japanese (JPN), Korean (KOR), Russian (RUS), Portuguese (PRT) and Spanish (ES). Audio signals were captured utilizing commercially available devices, specifically smartphones. These recordings were obtained from individuals who red designated scripts, contributing to a comprehensive and diverse collection of linguistic variations. Furthermore, for eight of the available accents, namely USA, CAN, GBR, CHN, KOR, PRT, RUS, and JPN, there is a provision of both a training dataset comprising approximately 20 hours and a corresponding  2 hours test set. In contrast, for the remaining two accents, CAN and ES, although no training data is available, an equivalent amount of testing data is provided. This distribution ensures a balanced  evaluation of the model's performance across the various accents. 

    \begin{table*}[ht]
        \centering
        \caption{ASR decoding results for the Interspeech 2020 test sets. Upper, light grey, ASR decoding results for real data only. $C$: model trained on Librispeech 960-hours \cite{LibriSpeech}. $FT$: model $C$ adapted to each accented training set separately. $FM$: model $C$ fine-tuned on the mix of all training accented sets. Lower, dark grey. ASR decoding results for TTS data augmentation (75\%/25\% TTS/real data). $FT_{aug}$, $FM_{aug}$ analogous as above with TTS augmentation. $FM_{aug+db}$: $FM_{aug}$ with batches built with 75\%/25\% TTS/real data.}
    
        \definecolor{Gray}{gray}{0.9}
        \definecolor{Gray_2}{gray}{0.75}
        
        \label{tab:Table 3}
        \begin{tabular}{cccccccccccc|c}
        \toprule
        \textbf{Model}&\multicolumn{11}{c}{\textbf{WER[\%]}} \tikz [remember picture] \node (rightmark) {}; \\
        \midrule
        &&GBR&USA&CAN&PRT&IND&CHN&ESP&KOR&RUS&JPN&Avg\\ 
        \cline{1-12}
        \rowcolor{Gray}
        \rowcolor{Gray}
        \tikz [remember picture] \node (n1) {}; 
        $\mathit{C}$ & & 7.47 & 17.74 & 18.26 & 19.28  & 23.40 & 24.03 & 26.35 & 33.78 & 35.75 & 36.97 & 24.3 \\
        \rowcolor{Gray}
        $\mathit{FT}$ & & 2.32 & 8.18 & X & 6.67 & 9.46 & 9.97 & X & 7.16 & 10.15 & 7.36 & 7.66 \\
        \rowcolor{Gray}
        $\mathit{FM}$ & & 2.06 & 4.95 & 4.51 & 5.31 & 8.51 & 9.87 & 7.39 & 5.40 & 8.98 & 6.31 & 6.33  \\
        \midrule
        TTS-AUG \\
        \midrule
        \rowcolor{Gray_2}
        $\mathit{FT_{aug}}$ & & 2.42 & 6.82 & X & 6.65 & 9.21 & 11.96 & X & 7.21 & 10.87 & 7.90 & 7.88 \\
        \rowcolor{Gray_2}
        $\mathit{FM_{aug}}$ & & 2.22 & 5.46 & 4.72 & 5.62 & 8.64 & 10.74 & 8.14 & 5.92 & 9.65 & 6.82 & 6.79\\
        \rowcolor{Gray_2}
        $\mathit{FM_{aug+db}}$ & & 2.25 & 5.38 & 4.63 & 5.50 & 8.50 & 11.21 & 8.00 & 5.85 & 9.47 & 6.87 & 6.77 \\   
        \bottomrule
        \end{tabular}
    \end{table*}

    \subsection{Single accent}
    
        \subsubsection{ASR fine-tuning on mixed real and TTS data}
        For preliminary results, we focused on one accent specifically, the Indian English accent. Initially, we trained the ASR model from scratch on the full 960 hours of \cite{LibriSpeech}. The model scored a WER of 2.34\% and 5.55\% on the \textit{test-clean} and \textit{test-others} partitions of Librispeech. However, when decoding the \textit{test-IND} partition of the Interspeech 2020 dataset with the aforementioned ASR model, we obtained a consistent performance degradation in terms of WER (22.24\%). This degradation in performance can be attributed to the inherent mismatch in acoustic and linguistic characteristics between the training data and the \textit{test-IND} partition, underscoring the need for further refinement and adaptation of the ASR model to effectively handle diverse and unseen conditions. \begin{table}[ht]
            \vspace*{+3pt}
    
            \centering
            \caption{ASR decoding results for the \textit{test-IND} set with ASR model fine-tuned on different partitions of \textit{train-IND} augmented with TTS generated data. R: model trained on real data. $\mathrm{R}-\mathrm{TTS_{c}}$: model trained on real data mixed with synthetic data generated from $\mathrm{TTS_{c}}$. $\mathrm{R}-\mathrm{TTS_{I}}$: model trained on real data mixed with synthetic data generated from $\mathrm{TTS_{I}}$. The last column refers to the WERR achieved by each augmentation strategy with respect to no augmentation.}
                
            \definecolor{Gray}{gray}{0.9}
            \definecolor{Gray_2}{gray}{0.75}
            
            \begin{tabular}{ccccc}
            \toprule
            \textbf{real data}&\multicolumn{3}{c}{\textbf{WER[\%]}}&\textbf{WERR[\%]} \tikz [remember picture] \node (rightmark) {}; \\
            \midrule
            &$\mathrm{R}$&$\mathrm{R}$-$\mathrm{TTS_{c}}$&$\mathrm{R}$-$\mathrm{TTS_{I}}$\\ 
            \midrule
            \rowcolor{Gray}
            
                    \rowcolor{Gray}
                    10\%    & 13.29  & 11.82 & 10.98  & 11/17\\
                    \rowcolor{Gray}
                    25\%    & 11.80  & 10.34 & 10.15  & 12/14\\
                    \rowcolor{Gray}
                    50\%    & 10.51  & 9.39  & 9.40   & 11/11\\
                    \rowcolor{Gray}
                    75\%    & 9.74   & 8.81  & 8.64   & 9.5/11\\
                    \rowcolor{Gray}
                    90\%    & 9.53   & 8.59  & 8.53   & 11/12\\
            
            \bottomrule
            \end{tabular}  
            \vspace*{01pt}
            \\ model trained on full accented set (14~hours): 8.33\% WER 
        \end{table} \\ Afterwards, we fine-tuned the ASR model on the full 20 hours of the \textit{train-IND} subset for 10 more epochs and decoded the \textit{test-IND} partition. We obtained a WER of 8.33\% which will be the lower-bound for the WER in the following TTS augmentation experiments on the Indian English accent. In the TTS augmentation experiments, we considered two ZS-TTS models: the first ($\mathrm{TTS_c}$, Table~2) is a pre-trained checkpoint provided by \cite{yourtts} which is trained on the LJ-Speech \cite{ljspeech17}, VCTK \cite{vctk}, the TTS-Portuguese Corpus \cite{Casanova_2022} and part of the M-AILABS dataset \cite{mailab}. The second model ($\mathrm{TTS_I}$, Table~2) is obtained from the previous checkpoint by continuing the training for 300 epochs on the three Indian speakers available in the VCTK dataset for a total of approximately 1.5~hours of speech divided equally across the speakers. To simulate different low resource scenarios, we sampled different fractions of the original \textit{train-IND} training dataset and augmenting it to the full 20 hours with TTS data generated from both of the aforementioned models. Importantly, the text input for the TTS model is specifically drawn from the original \textit{train-IND} set post-random sampling. This selection ensures the consistent preservation of textual content across all mixing conditions. This standardization is crucial for the accurate assessment of TTS augmentation and the systematic isolation of textual content, a factor that could potentially impact ASR performance, from the acoustic aspects under consideration. For example, to obtain the results for 25\% mixing in Table~2, we randomly selected 25\% of the \textit{train-IND} dataset and augmented it to 20 hours generating synthetic speech with both of the TTS models. Moreover, the synthetic speaker for each TTS utterance is randomly selected among the three available VCTK Indian speakers. Following the data generation process, we proceed to fine-tune the ASR model that was originally trained exclusively on the LibriSpeech 960-hour dataset. This fine-tuning is conducted across all datasets augmented using TTS. Subsequently, we perform decoding on the \textit{test-IND} subset using each fine-tuned model. The outcomes for each condition are presented in Table~2. Here, we can observe that WERs generated by the ASR models fine-tuned on augmented sets generated by the two TTS models ($\mathrm{TTS_{c}}$ and $\mathrm{TTS_{I}}$) are lower than using only real data ($\mathrm{R}$). However, the model fine-tuned on the Indian speakers from VCTK ($\mathrm{TTS_{I}}$) provides a higher WERR when compared with the publicly available checkpoint from \cite{yourtts}. 

        \subsubsection{ASR fine-tuning on TTS data}
    
        Furthermore, we fine-tuned the ASR model trained on the LibriSpeech 960-hour, exclusively on TTS-generated data. The results in Table~3 show that, in this condition, only the TTS model tuned on the three Indians speakers ($\mathrm{TTS_{I}}$) is able to provide a lower WER with respect to the original ASR model trained on Librispeech 960 (Dark grey rows, Table~3). However, this phenomenon is evident only when employing limited quantities of TTS data. We argue that this occurs because the ASR model exhibits overfitting tendencies when exposed to an excessive volume of TTS-generated data, resulting in a lapse of its ability to retain the speech characteristics inherent in real-world speech corpora.
        
        \begin{table}[ht]
        \vspace*{+3pt}

        \centering
        \caption{ASR decoding results for the \textit{test-IND} set with ASR model fine-tuned on TTS generated data. $\mathrm{TTS_{c}}$: model trained synthetic data generated from $\mathrm{TTS_{c}}$. $\mathrm{TTS_{I}}$: model trained synthetic data generated from $\mathrm{TTS_{I}}$. The last column refers to the WERR achieved by each augmentation strategy with respect to the model trained only on \cite{LibriSpeech}.}  
            
        \definecolor{Gray}{gray}{0.9}
        \definecolor{Gray_2}{gray}{0.75}
        
        \begin{tabular}{ccccc}
        \toprule
        \textbf{tts data}&\multicolumn{2}{c}{\textbf{WER[\%]}}&\textbf{WERR[\%]} \tikz [remember picture] \node (rightmark) {}; \\
        \midrule
        &$\mathrm{TTS_{c}}$&$\mathrm{TTS_{I}}$\\ 
        \midrule
        \rowcolor{Gray}

                \rowcolor{Gray_2}
                1.4 hours     & 23.67 & 21.13  & -6.4/5.0\\
                \rowcolor{Gray_2}
                3.5 hours     & 25.91 & 21.90  & -16.5/1.5\\
                \rowcolor{Gray}
                7 hours       & 28.44 & 22.5  &  -27.9/-1.2\\
                \rowcolor{Gray}
                10.5 hours    & 30.18 & 23.38   & -35.7/-5.1\\
                \rowcolor{Gray}
                12.6 hours    & 31.47 & 24.20  &  -41.5/-8.8\\
        \bottomrule
        
        \end{tabular}
        \vspace*{01pt}
        \\ model trained on real data only: 22.24\% WER
        \end{table}

    \subsection{Multiple accents}
    In this Section, we considered multiple accents without limiting to the Indian accented English case. In this context, we used the Interspeech 2020 accented English recognition challenge dataset \cite{shi2021accented} described in Section~3. For each accented partition, we selected the cleaner recordings according to the metadata annotations provided with the audio recording. However, this initial selection process resulted in imbalanced subset dimensions. To rectify this imbalance, we uniformly reduced the length of all partitions to match the duration of the shortest accented subset, amounting to approximately 11 hours, accomplished through a random sampling procedure. \\ Initially,  our baseline results were established using the ASR model (referred to as the $\mathit{C}$ model) trained on the complete 960-hour Librispeech dataset \cite{LibriSpeech}. First, we decoded each test accented subset with model $\mathit{C}$ (results shown in Table~1, first row). Second, we fine-tuned model $\mathit{C}$ on each accented training subset separately ($\mathit{FT}$, Table~1, second row). Finally, we fine-tuned the model $\mathit{C}$ by mixing all the training accented subsets together ($\mathit{FM}$, Table~1, third row). Decoding results for each accented test subset are reported in Table~1 (upper section, light grey rows). To train the TTS model to reproduce different accented voices, we used the same train subsets of the Interspeech 2020 accented English recognition challenge \cite{shi2021accented} described earlier. We started the training from the publicly available checkpoint \cite{yourtts} and continued for 300 further epochs. Furthermore, while training, the TTS model was guided to learn a different language embedding for each accent. For doing this, we modified the language embedding layers to fit the new accents (treated as new languages) and we randomly initialized the previously learnt weights for all language (now accent) embedding layers. \\ Following the results obtained for a single accent (Table~2, second row, best WERR), we mixed 75\% TTS-generated data with 25\% real data for all the multi-accent experiments (Table~1, lower section, dark grey rows). We repeated analogous ASR experiments as for the non-augmented sets. Results are presented in Table~3 (dark grey rows). In this case, it is seen that the TTS trained on multiple accented sets combined with accent embeddings does not improve ASR performance when used for data augmentation. This might be due to the TTS's inability to model all different pronunciation through accent embeddings or to the lower data quality of the Interspeech 2020 corpus. To test the latter hypothesis, we fine tuned $\mathrm{TTS_{c}}$ on three Indian speakers selected from the Interspeech 2020 dataset. \begin{table}[ht]
    
            \centering
            \caption{WER results of the \textit{test-IND} set. R-$\mathrm{TTS_{INI}}$: ASR model adapted on a mix of real and synthetic speech obtained from the TTS model fine-tuned on a small subset (3 speakers) of the \textit{train-IND} subset.}
                
            \definecolor{Gray}{gray}{0.9}
            \definecolor{Gray_2}{gray}{0.75}
            
            \begin{tabular}{cccc}
            \toprule
            \textbf{real data}&\multicolumn{2}{c}{\textbf{WER[\%]}} \tikz [remember picture] \node (rightmark) {}; \\
            \midrule
            &$\mathrm{R}$&$\mathrm{R}$-$\mathrm{TTS_{INI}}$\\ 
            \midrule
            \rowcolor{Gray}
            
                    \rowcolor{Gray}
                    10\%    & 13.29  & 11.68   \\
                    \rowcolor{Gray}
                    25\%    & 11.80  & 10.31   \\
                    \rowcolor{Gray}
                    50\%    & 10.51  & 9.55    \\
                    \rowcolor{Gray}
                    75\%    & 9.74   & 8.95    \\
                    \rowcolor{Gray}
                    90\%    & 9.53   & 8.86    \\
            
            \bottomrule
            \end{tabular}  

        \end{table} \\ The results for these experiments are given in Table~4 and show that this model produced augmentation results on par or worse than the pre-trained model (Table~2, $\mathrm{R}$-$\mathrm{TTS_{C}}$) from \cite{yourtts}. Because of this reason, we believe that, when compared with the VCTK speakers, the lower data quality of the Interspeech 2020 corpus severely impacts the quality of the synthetic accented speech used for augmentation.

\subsection{Combine TTS with other audio augmentations}

    Finally, we tested the effects of common speech augmentation techniques namely room impulse response (RIR), noise, speed and SpecAugment \cite{Park_2019} in comparison and in combination with TTS augmentation. During the training process, for each batch, we independently apply augmentations to all utterances. Subsequently, we concatenate the original and augmented samples to form the new batch. For RIR and noise augmentation, we used the OpenRIR \cite{Ko2017ASO} dataset which contains both real and simulated RIRs and noises. \begin{table}[ht]
   
            \centering
            \caption{WER results of the \textit{test-IND} set. $\mathrm{R_{aug}}$: ASR model fine-tuned on real data augmented with the techniques specified in Section~3.3. $\mathrm{R}-\mathrm{TTS_{INI}}$: ASR model fine-tuned on a mix of synthetic TTS speech obtained from the model adapted to a subset (3 speakers) of the \textit{train-IND} subset.} 
                
            \definecolor{Gray}{gray}{0.9}
            \definecolor{Gray_2}{gray}{0.75}
            
            \begin{tabular}{ccccc}
            \toprule
            \textbf{real}&\multicolumn{3}{c}{\textbf{WER[\%]}} \tikz [remember picture] \node (rightmark) {}; \\
            \midrule
            &$\mathrm{R}$&$\mathrm{R}_{aug}$&$\mathrm{R}_{aug}$-$\mathrm{TTS_{INI}}$ \\ 
            \midrule
            \rowcolor{Gray}
            
                    \rowcolor{Gray}
                    10\%    & 13.29  & 13.52  & 11.03 \\ 
                    \rowcolor{Gray}
                    25\%    & 11.80  & 11.90  & 10.08 \\ 
                    \rowcolor{Gray}
                    50\%    & 10.51  & 10.08  & 9.66  \\ 
                    \rowcolor{Gray}
                    75\%    & 9.74   & 9.33   & 9.10 \\ 
                    \rowcolor{Gray}
                    90\%    & 9.53   & 9.00   & 9.04 \\ 
            
            \bottomrule
            \end{tabular}  
        \end{table} \\ When dealing with accented speech, the findings presented in Table~5 indicate that relying solely on conventional speech augmentation techniques does not provide enhancements in ASR performance ($\mathrm{R}_{aug}$, Table~5). Moreover, the combination of the conventional speech augmentations with the synthetic TTS data, seems to harm the ASR. More precisely, in the case of $\mathrm{R_{aug}}$, conventional augmentation appears to degrade the ASR on smaller (10\% and 25\%) partitions but improves the WER for larger sets. However, in the case of $\mathrm{R}_{aug}$-$\mathrm{TTS_{INI}}$, we observe the opposite trend: limited improvements on smaller partitions and degradation on the larger (analogous results for model $\mathrm{TTS_{INI}}$).

\section{Conclusions}

    In this paper we describe a zero-shot TTS system application to the problem of biased ASR systems in low resource and under represented scenarios. We show that the pre-trained open-source model provides some WER improvement when used for data augmentation on a single accent. However, when the model is fine-tuned on a small amount of high-quality accented speech data the augmentation with synthetic data leads to statistically improved results when compared with the pre-trained model. Furthermore, when the ASR models are fine-tuned solely on TTS generated data, corresponding to the condition when no real accented speech is available, only the proposed method can improve the ASR WER. Moreover, we repurposed the language embeddings to encode accent specific pronunciations and fine-tuned the TTS model on the whole Interspeech 2020 dataset. However, this model didn't lead to improved WER when decoding real accented speech. We argue that this might be due to the fact that the Interspeech 2020 recordings are lower quality when compared to VCTK: this was tested by fine tuning the TTS model on a subset of the Indian training partition of the Interspeech dataset and using this model for ASR data augmentation. Finally we tested the effect of common speech augmentation techniques in combination with the TTS augmentation and show that this augmentation strategy harms the ASR when dealing with accented data. Further research will focus on understanding the fundamental difference between real and synthetic TTS speech from the perspective of different ASR models.

\section{Acknowledgements}
    This project was funded by the European Union’s Horizon 2020 program under the Marie Skłodowska-Curie grant No 956369.

\clearpage

\bibliographystyle{IEEEtran}
\bibliography{sapstrings,new_bib}

\begin{thebibliography}{10}
\providecommand{\url}[1]{#1}
\csname url@samestyle\endcsname
\providecommand{\newblock}{\relax}
\providecommand{\bibinfo}[2]{#2}
\providecommand{\BIBentrySTDinterwordspacing}{\spaceskip=0pt\relax}
\providecommand{\BIBentryALTinterwordstretchfactor}{4}
\providecommand{\BIBentryALTinterwordspacing}{\spaceskip=\fontdimen2\font plus
\BIBentryALTinterwordstretchfactor\fontdimen3\font minus \fontdimen4\font\relax}
\providecommand{\BIBforeignlanguage}[2]{{%
\expandafter\ifx\csname l@#1\endcsname\relax
\typeout{** WARNING: IEEEtran.bst: No hyphenation pattern has been}%
\typeout{** loaded for the language `#1'. Using the pattern for}%
\typeout{** the default language instead.}%
\else
\language=\csname l@#1\endcsname
\fi
#2}}
\providecommand{\BIBdecl}{\relax}
\BIBdecl

\bibitem{li2021accent}
J.~Li, V.~Manohar, P.~Chitkara, A.~Tjandra, M.~Picheny, F.~Zhang, X.~Zhang, and Y.~Saraf, ``Accent-robust automatic speech recognition using supervised and unsupervised wav2vec embeddings,'' \emph{arXiv preprint arXiv:2110.03520}, 2021.

\bibitem{DNN}
H.-J. Na and J.-S. Park, ``Accented speech recognition based on end-to-end domain adversarial training of neural networks,'' \emph{Applied Sciences}, vol.~11, no.~18, 2021.

\bibitem{f2n}
S.~Liu, D.~Wang, Y.~Cao, L.~Sun, X.~Wu, S.~Kang, Z.~Wu, X.~Liu, D.~Su, D.~Yu, and H.~Meng, ``End-to-end accent conversion without using native utterances,'' in \emph{ICASSP 2020 - 2020 IEEE International Conference on Acoustics, Speech and Signal Processing (ICASSP)}, 2020, pp. 6289--6293.

\bibitem{n2f}
\BIBentryALTinterwordspacing
P.~Klumpp, P.~Chitkara, L.~Sari, P.~Serai, J.~Wu, I.-E. Veliche, R.~Huang, and Q.~He, ``Synthetic cross-accent data augmentation for automatic speech recognition,'' \emph{ArXiv}, 2023. [Online]. Available: \url{https://arxiv.org/abs/2303.00802}
\BIBentrySTDinterwordspacing

\bibitem{yourtts}
E.~Casanova, J.~Weber, C.~D. Shulby, A.~C. Junior, E.~G{\"o}lge, and M.~A. Ponti, ``Yourtts: Towards zero-shot multi-speaker tts and zero-shot voice conversion for everyone,'' in \emph{{Int.} {Conf.} on Machine Learning}.\hskip 1em plus 0.5em minus 0.4em\relax PMLR, 2022, pp. 2709--2720.

\bibitem{speechbrain}
\BIBentryALTinterwordspacing
M.~Ravanelli, T.~Parcollet, P.~Plantinga, A.~Rouhe, S.~Cornell, L.~Lugosch, C.~Subakan, N.~Dawalatabad, A.~Heba, J.~Zhong, J.-C. Chou, S.-L. Yeh, S.-W. Fu, C.-F. Liao, E.~Rastorgueva, F.~Grondin, W.~Aris, H.~Na, Y.~Gao, R.~D. Mori, and Y.~Bengio, ``{SpeechBrain}: A general-purpose speech toolkit,'' 2021. [Online]. Available: \url{https://arxiv.org/abs/2106.04624}
\BIBentrySTDinterwordspacing

\bibitem{vits}
J.~Kim, J.~Kong, and J.~Son, ``Conditional variational autoencoder with adversarial learning for end-to-end text-to-speech,'' in \emph{Int. Conf. on Machine Learning}.\hskip 1em plus 0.5em minus 0.4em\relax PMLR, 2021, pp. 5530--5540.

\bibitem{hifi}
J.~Kong, J.~Kim, and J.~Bae, ``Hifi-gan: Generative adversarial networks for efficient and high fidelity speech synthesis,'' \emph{Advances in Neural Information Processing Systems}, vol.~33, pp. 17\,022--17\,033, 2020.

\bibitem{glowtts}
J.~Kim, S.~Kim, J.~Kong, and S.~Yoon, ``Glow-tts: A generative flow for text-to-speech via monotonic alignment search,'' \emph{Advances in Neural Information Processing Systems}, vol.~33, pp. 8067--8077, 2020.

\bibitem{HASP}
\BIBentryALTinterwordspacing
H.~S. Heo, B.-J. Lee, J.~Huh, and J.~S. Chung, ``Clova baseline system for the voxceleb speaker recognition challenge 2020,'' 2020. [Online]. Available: \url{https://arxiv.org/abs/2009.14153}
\BIBentrySTDinterwordspacing

\bibitem{CTC}
A.~Graves, S.~Fern{\'a}ndez, F.~Gomez, and J.~Schmidhuber, ``Connectionist temporal classification: labelling unsegmented sequence data with recurrent neural networks,'' in \emph{Proceedings of the 23rd {Int.} {Conf.} on Machine learning}, 2006, pp. 369--376.

\bibitem{vctk}
\BIBentryALTinterwordspacing
J.~Yamagishi, C.~Veaux, and K.~MacDonald, ``{CSTR VCTK Corpus}: English multi-speaker corpus for {CSTR} voice cloning toolkit,'' 2019. [Online]. Available: \url{https://datashare.ed.ac.uk/handle/10283/2950}
\BIBentrySTDinterwordspacing

\bibitem{shi2021accented}
\BIBentryALTinterwordspacing
X.~Shi, F.~Yu, Y.~Lu, Y.~Liang, Q.~Feng, D.~Wang, Y.~Qian, and L.~Xie, ``The accented english speech recognition challenge 2020: Open datasets, tracks, baselines, results and methods,'' \emph{ArXiv}, 2021. [Online]. Available: \url{https://arxiv.org/abs/2102.10233}
\BIBentrySTDinterwordspacing

\bibitem{LibriSpeech}
V.~Panayotov, G.~Chen, D.~Povey, and S.~Khudanpur, ``Librispeech: An asr corpus based on public domain audio books,'' in \emph{2015 IEEE {Int.} {Conf.} on Acoustics, Speech and Signal Processing (ICASSP)}, 2015, pp. 5206--5210.

\bibitem{ljspeech17}
\BIBentryALTinterwordspacing
K.~Ito and L.~Johnson, ``The {LJ} speech dataset,'' 2017. [Online]. Available: \url{https://keithito.com/LJ-Speech-Dataset/}
\BIBentrySTDinterwordspacing

\bibitem{Casanova_2022}
\BIBentryALTinterwordspacing
E.~Casanova, A.~C. Junior, C.~Shulby, F.~S. de~Oliveira, J.~P. Teixeira, M.~A. Ponti, and S.~Alu{\'{\i}}sio, ``{TTS}-portuguese corpus: a corpus for speech synthesis in brazilian portuguese,'' \emph{Language Resources and Evaluation}, vol.~56, no.~3, pp. 1043--1055, jan 2022. [Online]. Available: \url{https://doi.org/10.1007%2Fs10579-021-09570-4}
\BIBentrySTDinterwordspacing

\bibitem{mailab}
\BIBentryALTinterwordspacing
I.~Solak, ``The m-ailabs speech dataset.'' 2017. [Online]. Available: \url{https://www.caito.de/2019/01/the-m-ailabs-speechdataset/}
\BIBentrySTDinterwordspacing

\bibitem{Park_2019}
D.~S. Park, W.~Chan, Y.~Zhang, C.-C. Chiu, B.~Zoph, E.~D. Cubuk, and Q.~V. Le, ``Specaugment: A simple data augmentation method for automatic speech recognition,'' in \emph{Interspeech 2019}.\hskip 1em plus 0.5em minus 0.4em\relax ISCA, Sep. 2019.

\bibitem{Ko2017ASO}
T.~Ko, V.~Peddinti, D.~Povey, M.~L. Seltzer, and S.~Khudanpur, ``A study on data augmentation of reverberant speech for robust speech recognition,'' \emph{2017 IEEE International Conference on Acoustics, Speech and Signal Processing (ICASSP)}, pp. 5220--5224, 2017.

\end{thebibliography}

\end{document}